\documentclass[10pt,conference,letterpaper]{IEEEtran}
\usepackage{times,amsmath,epsfig}
\usepackage{graphicx,subfigure,caption}
\title{HelPal: A Search System for Mobile Crowd Service}
\author{\IEEEauthorblockN{Yao Wu, Tianzhen Wu, Ziyi Xiong, Yuncheng Wu, Hong Chen, Cuiping Li, Xiaoying Zhang}
\IEEEauthorblockA{Key Laboratory of Data Engineering and Knowledge Engineering of Ministry of Education, Beijing, China\\
School of Information, Renmin University of China, Beijing, China\\
Email: \{ideamaxwu, wutianzhen, xiong2613, yunchengwu, chong, licuiping, xyzruc\}@ruc.edu.cn}
}
\begin{document}
\maketitle
\begin{abstract} 
Proliferation of ubiquitous mobile devices makes location based services prevalent. Mobile users are able to volunteer as providers of specific services and in the meanwhile to search these services. For example, drivers may be interested in tracking available nearby users who are willing to help with motor repair or are willing to provide travel directions or first aid. With the diffusion of mobile users, it is necessary to provide scalable means of enabling such users to connect with other nearby users so that they can help each other with specific services. Motivated by these observations, we design and implement a general location based system HelPal for mobile users to provide and enjoy instant service, which is called mobile crowd service. In this demo, we introduce a mobile crowd service system featured with several novel techniques. We sketch the system architecture and illustrate scenarios via several cases. Demonstration shows the user-friendly search interface for users to conveniently find skilled and qualified nearby service providers.
\end{abstract}

%
\section{Introduction}
\begin{figure}
\centering
\includegraphics[width=1.0\columnwidth]{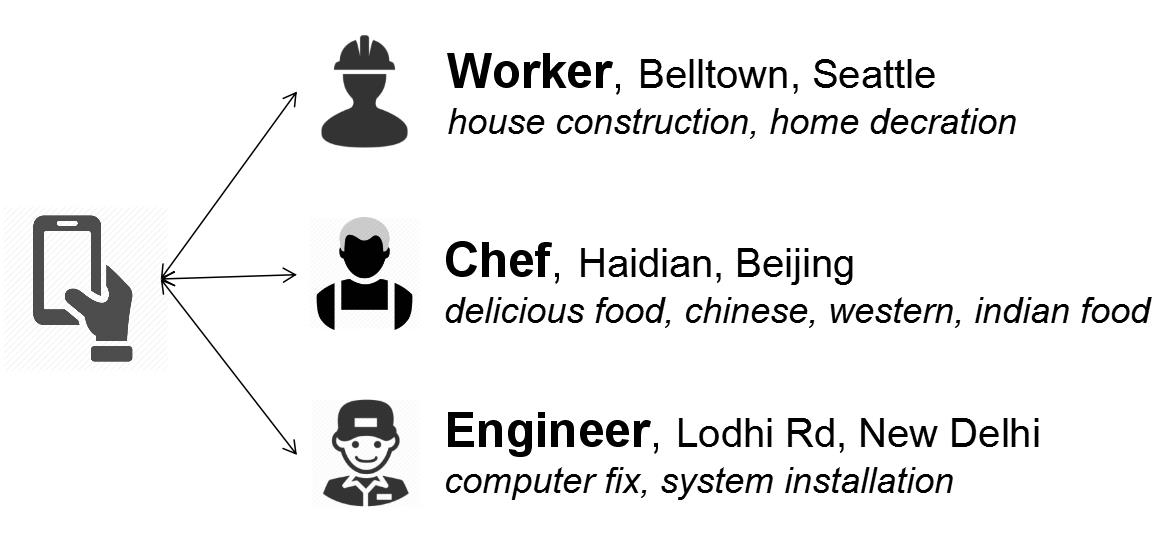}
\centering\caption{\label{ss}\footnotesize{Search for mobile crowd service provided by volunteered users with different skills of speciality.}}
\end{figure}

Popularity of mobile devices makes the location based services a compelling paradigm. Users carrying mobile devices can travel from places to places to collect various multimedia data, accomplish different tasks and provide qualified services.  Mobile users are able to volunteer as providers of specific services and in the meanwhile to search these services as shown in Fig. \ref{ss}. Sometimes we submit certain queries, but cannot get satisfying results from traditional web search engines, such as \emph{get me a cab}, \emph{fix my computer}, and \emph{I need a house cleaner right now}. From these queries, what we expect is some people who can provide specific services with domain skills. In reality, people with skills like taxi driving, computer fixing and house cleaning are just around us. It is necessary to connect these mobile users so that they can help each other with specific services.

Our system targets the increasing population of online mobile users, e.g., smartphone users, and enables such users to provide location-based services to each other. More specifically, on one hand, the system allows users to register as service volunteers, or micro-service providers. On the other hand, the system allows users to search available nearby relevant services provided by these volunteers.

\subsection{Motivating Examples}

Before all, we give an illustrated example of the scenarios where HelPal works. \emph{``Bob's computer doesn't work suddenly when he is preparing his report. He need his computer repaired as soon as possible to finish preparing his report before tomorrow. He can send query like `computer fixing' through HelPal. HelPal will return a ranked list of people who can do computer repair work near him. Then Bob can choose a qualified person to do it.''} In this example, the instant micro-service is expected. More examples can be given, such as drivers need available nearby users who are willing to help with motor repair or first aid when an accident happens. As we can see, results for such queries should satisfy domain skills, spatial proximity and timeliness.

\subsection{Existing Systems}

Location based services have been very successful, such as Yelp\footnote{https://www.yelp.com/}. On the contrary, the results of mobile crowd service are mobile users rather than static objects, e.g., restaurants or houses. Recently, Uber\footnote{https://www.uber.com/} thrives as a new kind of service, helping us find available drivers nearby. The limitation is that it can only provide certain service in vertical domain. Some special-interest and small-group needs are not well cared and satisfied.

Prevail of crowdsourcing motivates us to better utilize the power of grassroots. Until now, mobile crowdsourcing focuses on how to gather a collection of users to solve a macro project and it does not pay attention on the speciality of users. Take gMission\cite{ChenFZLXCCCTZ14} as an example, participants are not differentiated by their speciality. We cannot ask a doctor to fix a computer. Mobile crowd service focuses more on individual service and domain skills of users.

There are some web sites that help with the identification of trades people, e.g., Craigslist\footnote{https://www.craigslist.org/}. These tend not to be based on mobile devices, but they work off zip codes and the likes. Mobile crowd service is not only a more mobile, fuzzy-matching service. First, the traditional trades web sites provide physical commodity, i.e., used cars or electronics rather than individual service. Second, it is a post, repost and wait-for-answer model, which is not suitable for instant service. Our system is an end-to-end model with instant message module, enabling efficiency and timeliness.

\subsection{Challenges and Features}

Different from the existing systems, we design and implement a general location based service engine for mobile crowds to provide and enjoy instant micro-service. HelPal collects and organizes mobile users with their domain skills and updated locations by our client application and provides user-friendly interface for users to conveniently find the skilled and qualified people (turks\footnote{We name the qualified users who can provide specific service as turks after the Mechanical Turk, https://en.wikipedia.org/wiki/The\_Turk} as we call them) around to meet their needs. The system  features with two major aspects,
\begin{itemize}
\item Individual service with domain skills: It provides available and flexible service labor for users and any user can join to serve or be served with specific service, especially for the long-tail users;
\item Mobile and instant response: It takes the advantages of mobile devices to provide end-to-end service. It guarantees the time-efficiency of service query by indexing and other key techniques.
\end{itemize}

However, to implement such a mobile crowd service platform is not easy. The challenges lie in,
\begin{itemize}
\item Efficiency: The results of service query should be returned immediately, which needs efficient indexing technique for mobile spatial-textual objects (users);
\item Effectiveness: The results should return nearby qualified users just as the service query, which needs accurate spatial-textual matching techniques. 
\item Exploration: The results should also return potential turks that are qualified for tasks, which needs context-aware recommendation.
\end{itemize}


\section{System Overview}\label{sys}

\begin{figure}
\centering
\includegraphics[width=1.0\columnwidth]{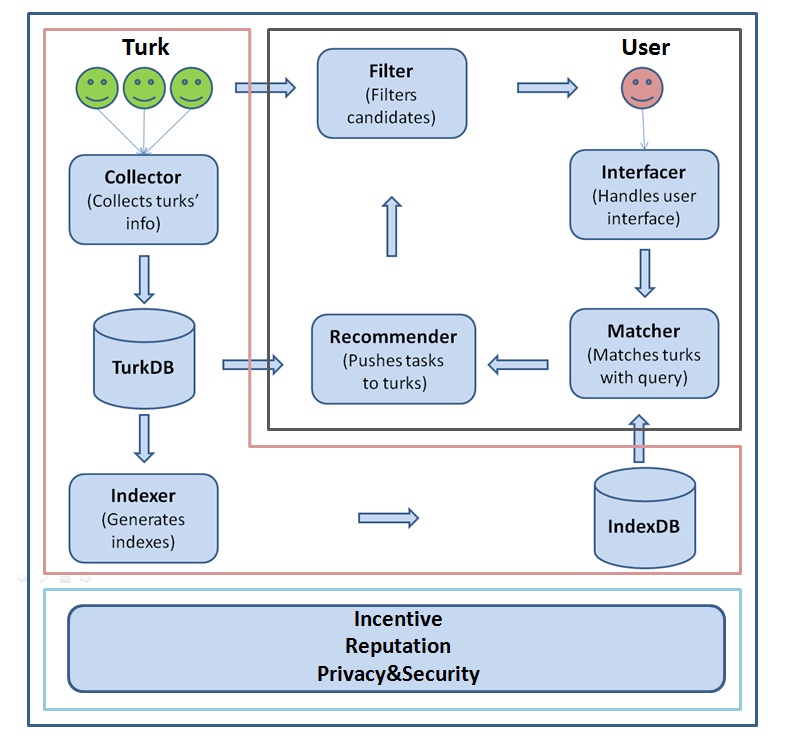}
\centering\caption{\label{arch}\footnotesize{Architecture of HelPal.}}
\end{figure}

In our system, users can be service requesters, service providers or both. In this paper, we use turks specifically to indicate the providers and users to refer to the requesters when it is not confusable. The architecture is mainly built based on indexing, matching and recommendation techniques to fulfill the efficiency and effectiveness as shown in Fig. \ref{arch}. In this section, we sketch our system via three core components, user management, service matcher and task recommendation. Incentive mechanism, reputation management and privacy protection are also concerned \cite{wu2016magicrowd} but skipped in this paper.

\subsection{User Management}
A turk is one of the grassroots who can provide service for users while a user is the one who requests for service. The turks are equipped with mobile devices that can access networks, sense locations, capture multimedia data. In HelPal, there are three key features to describe the turks, i.e., \emph{profile}, \emph{location} and \emph{action}. Profile is the basic information such as gender, interests and skills. The profile can describe what kinds of service the turks can provide. Location is the current location a turk senses, which determines the spatial range one can reach. Action is the daily log data about the turks, e.g., interactions with users, changes of profile and location. All the information about the turks is stored in \emph{TurkDB} with timestamp. The collector runs in background to keep record of turks when and after they register, which is kind of the crawler in search engine.

\noindent\textbf{Mobile Spatial Textual Objects Indexing}

\emph{Definition1: Spatial-Textual Object.} A spatial-textual object is represented with a triple $o=\langle\psi,l,t\rangle$, where $o.\psi$ is the set of skills, $o.l$ is the latitude and longitude of the turk's location and $o.t$ is the last positioned time.

\begin{figure}
\centering
\subfigure[]{
\includegraphics[width=0.9\columnwidth]{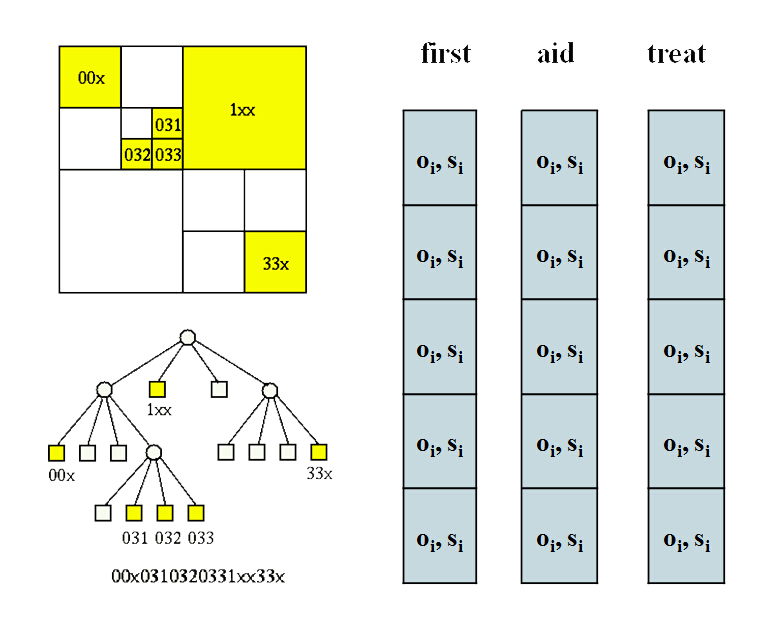}}
\subfigure[]{
\includegraphics[width=1.0\columnwidth]{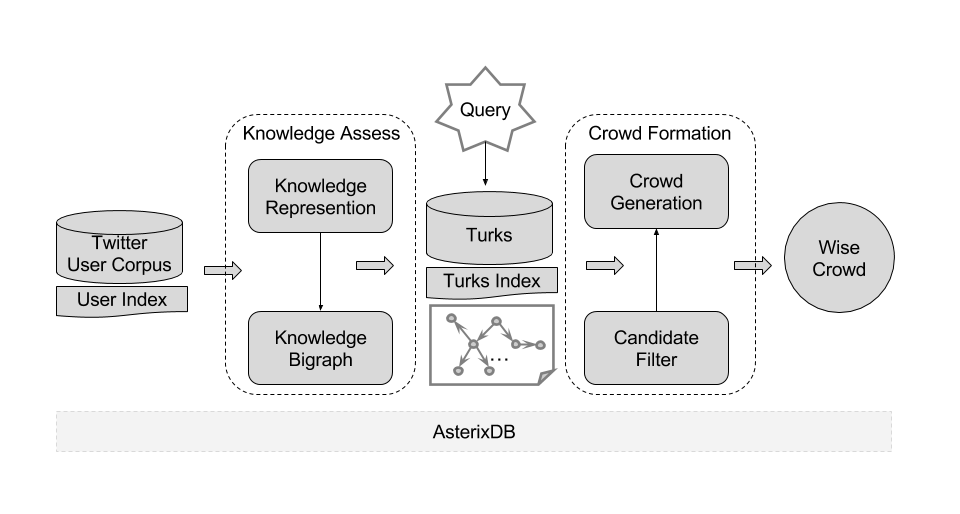}}
\subfigure[]{
\includegraphics[width=0.9\columnwidth]{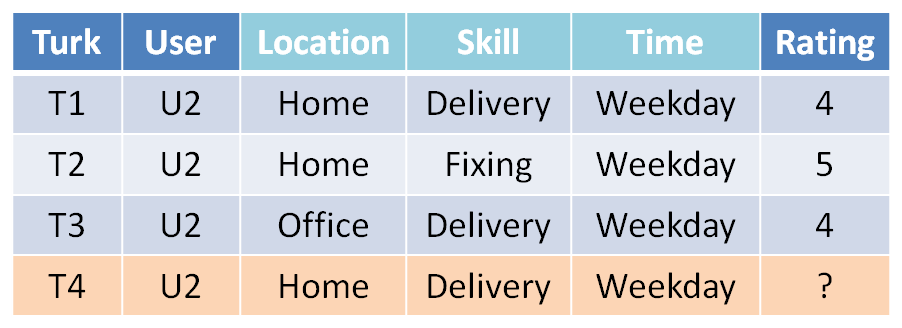}}
\centering\caption{\label{index}\footnotesize{(a) The storage structure of BIG-tree. (b) Framework of distant-supervision domain knowledge matching. (c) An example of context-aware recommendation.}}
\end{figure}

In this demo, there are two main challenges to handle users: 1) both spatial and textual information should be concerned. 2) the mobility of the users makes it difficult to construct and update the index for efficiency. Existing geo-textual indices do not consider the mobile information of data and cannot deal with the challenges. They are designed for processing spatial-keyword queries on static geo-textual objects. We focus more on the mobility rather than the motion perspective, e.g., direction and speed. We model the mobile information as probabilistic instances and present a new hybrid indexing for the mobile spatial textual objects, called BIG-tree \cite{Yaoaxe17} as shown in Fig. \ref{index} (a). The indexes are based on quadtree with sorted inverted lists and stored in \emph{IndexDB}. And we propose an improved threshold algorithm with lazy refinement and prior termination to efficiently process the Top-$k$ query over a large number of mobile spatial textual objects based on the index.

\subsection{Service Matcher}
The service can be expressed by various kinds of skills in different locations. Two features need to be considered in service matching, \emph{different skills} and \emph{spatial locations}. In most literature of moving objects query research, they only focus on the spatial similarity without consideration of textual information. However, in most location based service, we query objects not only with spatial approximation but also textual similarity. In consideration of mobility of users, we introduce time recency \cite{ChenCCT15} to infer the validity of spatial location. Service matcher computes the matching score for each candidate turk with their skills and locations.

\noindent\textbf{Temporal Spatial Textual Objects Matching}

\emph{Definition2: Temporal Spatial-Textual Query.} A service query $q=\langle\psi,l,t\rangle$, where $q.\psi$ is the set of query keywords, $q.l$ is the query location and $q.t$ is the query time.

The temporal spatial-keyword score of a spatial-textual object $o$ at time $t$ is defined as follows,
\begin{equation}\label{match}
    S(q,o)=[\alpha S_l(q.l,o.l)\!+\!(1\!-\!\alpha)S_\psi(q.\psi,o.\psi)] S_t(q.t,o.t)
\end{equation}
where $S_l(q.l,o.l)$ is the \emph{spatial proximity score} of distance between query $q$ and object $o$, $S_\psi(q.\psi,o.\psi)$ indicates \emph{the text relevance} between $q$ and object $o$, and $S_t(q.t,o.t)=\lambda^{-(q.t-o.t)}$ is the time recency. $\lambda$ is the base number that determines the rate of the recency decay. Time recency is a key factor measuring the spatial importance based on locations updated by mobile objects. The validity of a location decreases when it is not updated for a long time. To measure the text relevance, we propose PIN \cite{Yaopin17} as shown in Fig. \ref{index} (b), a domain knowledge matching with distant supervision to overcome the mismatch problem which is only based on keywords. We use Google Knowledge Graph as the distant supervision and model the expertise of each user as a knowledge bi-graph.

\subsection{Task Recommendation}
Content based matching is an accurate matching, which may miss some potential qualified turks. Besides, turks sometimes cannot update their skills timely or accurately. To tackle these problems, recommendation techniques can be applied to discover the potential qualified turks. Given the query, we use context-aware recommendation technique to find the potential turks and push the tasks to them, waiting their responds.

\noindent\textbf{Context-aware Recommender with Explicit Ratings}

\emph{Context-aware recommender system (CARS)} \cite{AdomaviciusT11} deals with modeling and predicting user tastes and preferences by incorporating available contextual information into the recommendation process as additional categories of data. These long-term preferences and tastes are usually expressed as ratings and are modeled as the function of not only items and users, but also of the context as shown in Fig. \ref{index} (c). In other words, ratings are defined with the rating function as
\begin{equation}\label{reco}
    R:User\times Turk\times Context \to Rating
\end{equation}
where \emph{User} and \emph{Turk} are the domains of users and turks respectively, \emph{Rating} is the domain of ratings, i.e., the satisfaction of users for the service of turks. \emph{Context} specifies the contextual information associated with the application, which are time, location and domain skills information in our model. In our demo, we incorporate the explicit information, e.g., ratings for the completed service, to recommend the service tasks for turks \cite{Yaorock17}.

After the matching and recommendation score computation, we return top-$k$ turks to users. After  the task is pushed to the candidates, they can choose to serve or not, which is not like the static web pages. When the turks from both matcher and recommender get tasks pushed from the server, they can accept, refuse or ignore. And this affects the candidate list. Filter is a re-ranking process based on the turks' choices. This also leads to some non-candidates becoming candidates due to the quit of some candidates.


\section{Demonstration}\label{dem}
\begin{figure}
  \centering
  \subfigure[]{
    \includegraphics[width=0.45\columnwidth]{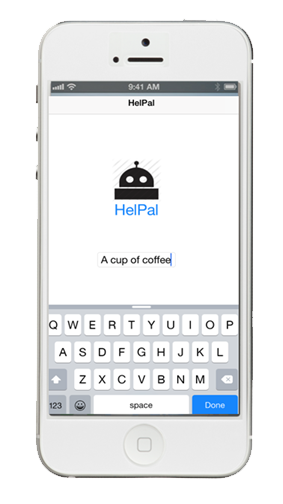}}
  \subfigure[]{
    \includegraphics[width=0.45\columnwidth]{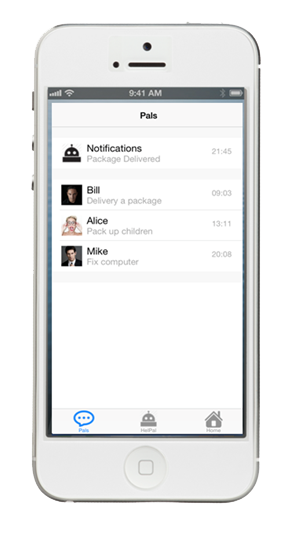}}
  \subfigure[]{
    \includegraphics[width=0.45\columnwidth]{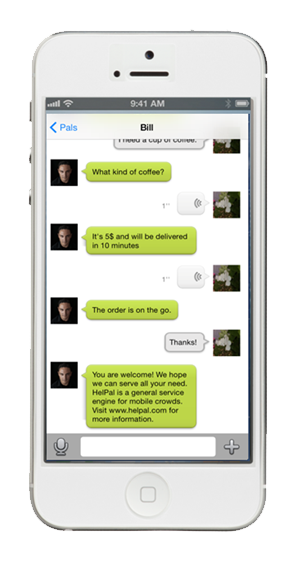}}
  \subfigure[]{
    \includegraphics[width=0.45\columnwidth]{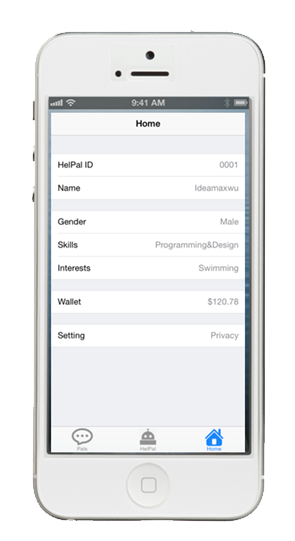}}
  \caption{\label{demo}\footnotesize{Demonstration of HelPal.}}
\end{figure}

A brief demonstration of HelPal is shown in Fig. \ref{demo}. It is necessary for users to register in HelPal and update their skills and location information before using it. \textbf{Service query (a)} is the interface of query submission. Users can submit their service queries into the query box. \textbf{Service notification (b)} handles notification of the task recommendation to turks and candidate turks to users. This can guarantee the timeliness of the request. \textbf{Service message (c)} manages the back-and-forth communication between users and turks. The details of service can be explained clearly through instant messages. \textbf{Profile panel (d)} is the information about users and turks and the profile is submitted by users and turks when they register or later in use. To be noted, most components and techniques run in background and we do not show it explicitly. 

HelPal is implemented by proposed novel search and recommendation techniques with succinct interfaces to fulfill efficiency and effectiveness to provide instant mobile crowd services, which is initially proposed for mobile computing and crowdsourcing communities. HelPal is further designed to connect the world-wide available human service to form an elastic and on-demand service pool. For demonstration, we need a laptop to run the server and a mobile device (e.g., iPhone) to run the application. Both the server and client side need wireless network access. A demo video can be found at https://youtu.be/fjGurUo4UbU.

\small
\bibliographystyle{unsrt}
\bibliography{refs}  

\end{document}